\documentclass[twocolumn]{aastex63}
\usepackage{amsmath}
\usepackage{color}
\usepackage{ulem}
\usepackage{graphicx}
\usepackage{textcomp}
\usepackage[T1]{fontenc}
\usepackage{footnote}

\begin{document}

\title{Light curves and polarizations of gravitationally lensed kilonovae}

\author[0000-0002-1768-0773]{Yan-Qing Qi}
\affiliation{Department of Astronomy, Xiamen University, Xiamen, Fujian 361005, China; tongliu@xmu.edu.cn}
\author[0000-0001-8678-6291]{Tong Liu}
\affiliation{Department of Astronomy, Xiamen University, Xiamen, Fujian 361005, China; tongliu@xmu.edu.cn}

\begin{abstract}
Kilonovae are generally believed to originate from the ejecta of binary neutron stars (NSs) or black hole and NS mergers. Free neutrons might be retained in the outermost layer of the ejecta to produce a precursor via $\beta$-decay. During the propagation of kilonovae to observers, a small percentage of them might be gravitationally lensed by foreground objects. In this paper, three lens models, i.e., the point-mass model, the singular isothermal sphere (SIS) model, and the Chang-Refsdal model, were taken into consideration to explore the light curves and polarizations of gravitationally lensed kilonovae. We found that if the time delay between two images exceeds the ejecta heating timescale for the lens mass $\sim 10^{10}~M_\odot$ in the SIS model, a tiny bump-like signal will be generated in the light curve, and the total luminosity will be magnified in all cases. The polarization of lensed kilonovae is significantly enhanced in most cases. Future detections of lensed kilonovae will impose constraints on the morphology of the ejecta and aid in the determination of the nature of compact object mergers and the search for strong gravitational lenses.
\end{abstract}

\keywords{polarimetry (1278), strong gravitational lensing (1643), transient sources (1851)}

\section{Introduction}

Following GW 170817 from a binary neutron star (NS) merger \citep{Abbott2017a}, a faint short gamma-ray burst (GRB) GRB 170817A \citep[e.g.,][]{Abbott2017b,Goldstein2017,Savchenko2017}, X-ray and radio afterglows \citep[e.g.,][]{Alexander2017,Haggard2017,Hallinan2017,Kim2017,Margutti2017,Troja2017}, and an ultraviolet-optical-infrared transient AT 2017gfo in the galaxy NGC 4993 were detected \citep[e.g.,][]{Arcavi2017,Chornock2017,Coulter2017,Cowperthwaite2017,Drout2017,Evans2017,Kasliwal2017,Pian2017,Smartt2017,Soares-Santos2017,Tanvir2017}. At 1.46 days after that merger, \citet{Covino2017} detected the null linear optical polarisation of AT 2017gfo at a degree of $(0.50\pm0.07)\%$. AT 2017gfo, as the first kilonova electromagnetic counterpart, offers us an excellent opportunity to examine the nucleosynthesis of $r$-process elements.

The kilonova corresponding to the merger of binary NSs or a black hole (BH) and an NS is thought to be powered by the radiative decay heating of $r$-process elements nucleosynthesized in the neutron-rich ejecta \citep[e.g.,][]{Li1998,Metzger2010,Metzger2019}. Multicomponent ejecta, such as the ``blue'' and ``red'' components from optical ($\sim$ several days) to NIR emissions ($\sim$ several weeks), were introduced to explain the kilonova emission \citep[e.g.,][]{Drout2017,Kasen2017,Kasliwal2017,Perego2017,Tanaka2017,Zhu2020}. The polar material dynamically squeezed by two merging NSs \citep[e.g.,][]{Oechslin2006,Wanajo2014,Radice2016,Sekiguchi2016}, the neutrino irradiation from the remnant hypermassive NS \citep[e.g.,][]{Metzger2014,Perego2014,Yu2018}, or outflows from the postmerger remnant BH disk \citep[e.g.,][]{Fernandez2013,Just2015,Siegel2017,Song2018,Fernandez2019,Fujibayashi2020,Qi2022} is linked to the lanthanide-free ``blue'' component. The tidally stripped equatorial material is associated with the lanthanide-rich ``red'' component \citep[e.g.,][]{Rosswog1999,Foucart2014,Kyutoku2015,Kyutoku2018,Chornock2017}. \citet[]{Bulla2019} and \citet{Bulla2021} found that polarization originated from the $r$-process ejecta can reach a detectable level at one to two days after the merger. A fast expanding layer known as the fast velocity tail is ejected first and located at the head of the $r$-process ejecta \citep[e.g.,][]{Kyutoku2014,Ishii2018,Radice2018a,Radice2018b}. The fast tail has a short expansion timescale during which the neutron-capture reaction may not proceed efficiently, leaving free neutrons \citep{Metzger2015}, which produce bright emission known as neutron precursors and exhibit a degree of polarization in the first hour after merger \citep[e.g.,][]{Matsumoto2018,Li2019}.

Thomson scattering by free electrons (the product of the neutron $\beta$-decay) can contribute significantly to opacity in kilonovae at an early stage \citep[e.g.,][]{Tanaka2018,Banerjee2020} and linearly polarize the escaping radiation. The geometry and composition of the ejected material, as well as the interaction of the various ejecta components, are all factors that might affect the polarization signals that a distant observer receives. \citet{Matsumoto2018} and \citet[][]{Li2019}, focusing on the electron-scattering dominated period in the first hour after the merger, anticipated a polarization signal produced by free neutrons up to $3\%$ for binary NS and BH-NS mergers, respectively. \citet{Bulla2019} used a Monte Carlo simulation to examine the polarization produced by $r$-process ejecta in the binary NS merger scenario that were comparable to those of AT 2017gfo. Early polarization observations can be used to estimate the mass of free neutrons, the nucleosynthesis conditions of the $r$-process, and the early radiation mechanism.

There have been several gravitational wave (GW) alerts released for either binary NS or BH-NS mergers during the observational run of LIGO/Virgo collaboration (LVC). However, despite numerous efforts to identify confirmed electromagnetic counterpart candidates, except for AT 2017gfo, no confirmed candidate has been identified thus far \citep[e.g.,][]{Anand2021,Andreoni2020,Coughlin2020,Page2020,Kasliwal2020,Sagues2021}. One probable explanation for the lack of electromagnetic counterparts is that they are inherently absent. For instance, in BH-NS mergers, where the NS is swallowed whole by the BH (the so-called ``plunging'' event) rather than being tidally disrupted, there may not be any material left outside of the merged BH remnant \citep[e.g.,][]{Zhu2021a}. However, it is also possible that current electromagnetic counterpart searches are too shallow to obtain distance and volumetric coverage for LVC probability maps \citep[e.g.,][]{Coughlin2020,Sagues2021,Zhu2021a}. That is, if kilonovae have low brightness and are separated by long distances, the kilonovae may be too faint to be detected \citep[e.g.,][]{Zhu2021b,Qi2022}. A strong gravitational lens provides an additional way to increase the detectable kilonovae brightness. As pointed out in several studies, a fraction of GRBs is probably gravitationally lensed by foreground objects during their propagation \citep[e.g.,][]{Paczynski1986,Mao1992,Wambsganss1993,Grossman1994,Nowak1994,Nemiroff1995,Paynter2021,Wang2021,Yang2021,Chen2022,Gao2022}.

In this work, we focus on the lensed kilonovae powered by the $r$-process element and free neutron radioactive decay heating. The paper is organized as follows. In Section 2, we first review the fundamental features of the rapidly expanding ejecta and the associated kilonova light curves. We then describe how to calculate the degree of polarization produced in kilonovae. In Section 3, we introduce the magnification and time delay of three different gravitational lens models: the point-mass (PM) model, the singular isothermal sphere (SIS) model, and the Chang-Refsdal (CR) model. In Section 4, we present the light curves and polarization results of the lensed kilonovae. A brief summary is given in Section 5.

\section{Kilonova Model}
\subsection{Light curves}\label{KN light curves}

It is anticipated that binary NS and BH-NS mergers will release neutron-rich ejecta. Here, we consider that the ejecta is composed of red and blue components with a fast velocity tail in the outermost layer of both. We assume that the half-opening angle of the lanthanide-rich red component in the equatorial region is $30^{\circ}$, and the rest of the remaining area is the lanthanide-free blue component \citep{Bulla2019}. The red or blue component, as a basic component, has a power-law density profile that can be expressed as \citep[e.g.,][]{Hotokezaka2013,Nagakura2014,Matsumoto2018}
\begin{equation}
\begin{array}{l}
\rho_{\rm {base}}(v, t,\theta)=\frac{M_{\rm {base}}}{2 \pi\left(v_{\rm {b}} t\right)^{3}}\left(\frac{v}{v_{\rm {b}}}\right)^{-\beta_{\rm {b}}}\left[\frac{\beta_{\rm {b}}-3}{1-\left(\frac{v_{\rm {t}}}{v_{\rm {b}}}\right)^{3-\beta_{\rm {b}}}}\right], v_{\rm {b}}<v<v_{\rm {t}},
\end{array}
\end{equation}
where $M_{\rm {base}}$, $\beta_{\rm {b}}=4$, $v_{\rm {b}}=0.1~c$, and $v_{\rm {t}}=0.3~c$ are the mass of red or blue component, the power-law index, the lowest velocity of the basic component, and the lowest velocity of the fast tail. We set $M_{\rm base}=0.03~M_{\odot}$ and $0.02~M_{\odot}$ for the fiducial values, respectively \citep[for a review see][]{Metzger2019}.

The ejecta is close to homologous expansion after mergers, so the fast tail moving ahead in the $r$-process ejecta has the density profile \citep[e.g.,][]{Kyutoku2014,Hotokezaka2018}
\begin{equation}
\begin{array}{l}
\rho_{\text {tail}}(v, t)=\frac{M_{\mathrm{n}}}{4 \pi\left(v_{\mathrm{t}} t\right)^{3}}\left(\frac{v}{v_{\mathrm{t}}}\right)^{-\beta_{\text {t }}}\left[\frac{\beta_{\text {t}}-3}{1-\left(\frac{c}{v_{\mathrm{t}}}\right)^{3-\beta_{\text {t}}}}\right], v_{\text {t}}<v,
\end{array}
\end{equation}
where $M_{n}=10^{-4}~M_{\odot}$, and $\beta_{\rm {t }}=6$ are the total tail mass (see below, for the composition of the tail), and the power-law index of the fast tail, respectively.

Since the material distribution is anisotropic, we calculate the material distribution of a unit solid angle. The mass of matter outside the $v$-shell is
\begin{equation}
M(>v)=\int_{v t}^{c t} \left(\rho_{\text {base}}+\rho_{\text {tail}}\right)r^{2}\sin \theta \mathrm{d} \theta \mathrm{d} \varphi \mathrm{d}r.
\end{equation}

Regarding the chemical composition of the different layers, we assume that the fastest layer ($v>v_{\rm n}$, $v_{\rm {n }}=0.5~c$ is the lowest velocity of the free neutrons) contains only free neutrons, i.e., $X_{\mathrm{r}}(v, t)=0$, and the red or blue component ($v_{\rm {b }}<v<v_{\rm {t }}$) contains only the $r$-process materials, i.e., $X_{\mathrm{r}}(v, t)=1$. For the inner part of the high-velocity tail component ($v_{\rm {t }}<v<v_{\rm {n }}$), the mass fractions of the $r$-process elements, free neutrons and protons (the products of the $\beta$-decay of neutrons) are
\begin{equation}
X_{\mathrm{r}}(v, t)=\frac{2}{\pi} \arctan \left[\left(\frac{M(>v)}{M_{\mathrm{n}}}\right)^{\alpha}\right],
\end{equation}
\begin{equation}
X_{\mathrm{n}}(v, t)=\left(1-X_{\mathrm{r}}\right) e^{-t / t_{\mathrm{n}}},
\end{equation}
\begin{equation}
X_{\mathrm{p}}(v, t)=\left(1-X_{\mathrm{r}}\right)\left(1-e^{-t / t_{\mathrm{n}}}\right),
\end{equation}
respectively, where the index $\alpha=10$ represents how drastically the nucleon abundance decreases in the $v<v_{\rm n}$ range, and $t_{\rm n}=900~\rm s$ is the $\beta$-decay timescale for neutrons.

The opacity in the ejecta arises from the electron scattering and the bound-bound transition of the $r$-process elements, so the optical depth is
\begin{equation}\label{tau}
\tau(v, t)=\int_{v t}^{c t} \mathrm{~d} r\left[\kappa_{\mathrm{r}} \rho_{\mathrm{base}}+\left(\kappa_{\mathrm{r}} X_{\mathrm{r}}+\kappa_{\mathrm{es}, \mathrm{H}} X_{\mathrm{p}}\right) \rho_{\text {tail}}\right],
\end{equation}
where $\kappa_{\mathrm{r,red}}=10~\rm cm^{2}~g^{-1}$ and $\kappa_{\mathrm{r,blue}}=1~\rm cm^{2}~g^{-1}$ are the bound-bound opacities of the red and blue components, respectively, and $\kappa_{\mathrm{es,H}}=0.04~\rm cm^{2}~g^{-1}$ is the Thomson scattering opacity of hydrogen. In Equation (\ref{tau}), we neglect the contribution of electrons produced by ionization of $r$-process elements because of the low opacity. We solve for the diffusion layer velocity $v_{\rm d}$ and the photospheric velocity $v_{\rm ph}$ at each time with $\tau=c/v$ and $\tau=1$, respectively.

The bolometric luminosity is provided by the total mass of $r$-process elements and free neutrons contained outside the diffusion layer, which are given, respectively, by
\begin{equation}
M_{\mathrm{r}}\left(>v_{\text {\rm d}}\right)=\int_{v_{\rm d}t}^{c t} \left(\rho_{\text {base}}+X_{\mathrm{r}} \rho_{\text {tail}}\right)r^{2} \sin \theta \mathrm{d} \theta \mathrm{d} \varphi \mathrm{d}r,
\end{equation}
and
\begin{equation}
M_{\mathrm{n}}\left(>v_{\text {\rm d}}\right)=\int_{v_{\rm d}t}^{c t} X_{\mathrm{n}} \rho_{\text {tail}}r^{2} \sin \theta \mathrm{d} \theta \mathrm{d} \varphi \mathrm{d}r.
\end{equation}

The bolometric luminosity is approximately calculated by
\begin{equation}
L \simeq q_{\mathrm{r}} M_{\rm r}\left(>v_{\mathrm{d}}\right)+q_{\mathrm{n}} M_{\mathrm{n}}\left(>v_{\mathrm{d}}\right),
\end{equation}
where $q_{\mathrm{r}}$ and $q_{\mathrm{n}}$ are the particular heating rates of the radiative decay from $r$-process elements and free neutrons, which are given by \citet{Wanajo2014} and \citet{Kulkarni2005}, respectively, as shown below:
\begin{equation}
q_{\mathrm{r}}=2 \times 10^{10}\left(\frac{t}{\text {day}}\right)^{-1.3} \mathrm{erg}~\mathrm{s}^{-1} \mathrm{~g}^{-1},
\end{equation}
and
\begin{equation}
q_{\mathrm{n}}=3 \times 10^{14}~ \mathrm{erg}~\mathrm{s}^{-1} \mathrm{~g}^{-1}.
\end{equation}

The effective temperature at the radius of the photosphere, i.e., $R_{\rm ph}=v_{\rm ph}t$, is
\begin{equation}
T_{\mathrm{eff}}=\left(\frac{L}{\sigma_{\mathrm{SB}} \sin \theta \mathrm{d} \theta \mathrm{d} \varphi  R_{\mathrm{ph}}^{2}}\right)^{1 / 4},
\end{equation}
where $\sigma_{\mathrm{SB}}$ is the Stefan-Boltzmann constant.

The observed flux per unit solid angle is
\begin{equation}
d F_{\nu}\left(\nu, t_{\mathrm{obs}}\right)=\frac{2 \pi h \nu^{3}}{c^{2}} \frac{1}{1-\exp \left(h \nu / k_{\mathrm{B}} T_{\mathrm{eff}}\right)} \frac{R_{\mathrm{ph}}^{2} \sin \theta \mathrm{d} \theta \mathrm{d} \varphi}{4 \pi D_L^{2}},
\end{equation}
where $h$, $k_{\rm B}$ and $D_L$ are the Planck constant, Boltzmann constant and luminosity distance of the source, respectively.

The observation time is
\begin{equation}
t_{\mathrm{obs}}=t+\frac{\left[R_{\mathrm{ph}}\left(\theta_{\mathrm{obs}}\right)-R_{\mathrm{ph}}(\theta)\right] \cos \alpha}{c},
\end{equation}
where $t$ is the emission time of the photons and $\cos \alpha=\sin\theta\cos\varphi\sin\theta_{\rm obs}\cos\varphi_{\rm obs}+\sin\theta\sin\varphi\sin\theta_{\rm obs}\sin\varphi_{\rm obs}+\cos\theta\cos\theta_{\rm obs}$ is the direction cosine between the direction of a unit solid angle at the photosphere radius and the line of sight.

If we assume that the observer is located at $\varphi_{\rm obs}=0^{\circ}$, the range of longitudes visible to an observer located in a certain $\theta$ is $[-\varphi(\theta),\varphi(\theta)]$, where $\varphi(\theta)$, as the half-day arc, is given by
\begin{equation}
\varphi(\theta)=\arccos \left(-\cot \theta \cot \theta_{\mathrm{obs}}\right).
\end{equation}

If the kilonova emission is horizontal axisymmetry, the total flux of that reaching the observer at $\theta_{\rm obs}$ is written as
\begin{equation}
F_{\nu}\left(\nu, t_{\mathrm{obs}}\right)=2 \int_{0}^{\pi} \int_{0}^{\varphi(\theta)} d F_{\nu}.
\end{equation}

Finally, we convert $F_{\nu}\left(\nu, t_{\mathrm{obs}}\right)$ to a monochromatic AB magnitude by the following equation:
\begin{equation}
M_{\nu}=-2.5 \log _{10}\left(F_{\nu} / 3631 \mathrm{Jy}\right).
\end{equation}

\subsection{Polarizations}

Extensive research has been conducted to estimate the degree of polarization in nonspherical supernovae explosions. The polarization of oblate or prolate spheroidal atmospheres was specifically computed analytically by \citet[][]{Shapiro1982} as a function of the asphericity parameter. \citet{Hoflich1991} and \citet{Kasen2003} used a different Monte Carlo method to calculate photon propagation.

We consider that a kilonova consists of two components, red and blue, with a fast tail in the outermost layer. When the photosphere is located in the fast velocity tail, electron scattering dominates the opacity and causes a large degree of polarization. When the photosphere recedes into the $r$-process element-rich ejecta, the opacity is dominated by the bound-bound transition, which does not result in polarization, as opposed to electron scattering \citep[e.g.,][]{Kasen2013,Tanaka2013}. Of course, the ionization of the $r$-process elements can supply electrons, the opacity of which is much lower than that of the nucleon ejecta in our calculation \citep[e.g.,][]{Kasen2013,Kyutoku2015}. \citet{Bulla2019} found that electron scattering opacities are comparable to the bound-bound opacities in the lanthanide-free component at R band and 1.5 days after the merger in their simulations.

\citet{Brown1977} derived an analytic expression of polarization for optically thin Thomson scattering in an axisymmetric envelope. Optically thin condition is a critical approximation for avoiding multiple scattering; and the polarization level in the optically thin case is proportional to the optical depth for electron scattering. The net polarization is as follows \citep[e.g.,][]{Brown2000,Friend1986}:

\begin{equation}\label{polarization}
P=\frac{3}{16} \sin ^{2} \theta_{\rm obs} \int_{0}^{\pi} \tau_{\mathrm{es}}(\theta) \left(1-3 \cos^{2}\theta\right) \sin\theta d\theta,
\end{equation}
where $\tau_{\mathrm{es}}$ is the electron scattering opacity at the photosphere, which is calculated as
\begin{equation}
\tau_{\mathrm{es}}(\theta)= \int_{v_{\rm ph}t}^{c t}\kappa_{\mathrm{es},\mathrm{H}}\left[\frac{x}{A}\left(\rho_{\text {base }}+X_{\mathrm{r}} \rho_{\text {tail }}\right)+X_{\mathrm{p}} \rho_{\text {tail }}\right]dr,
\end{equation}
including the contributions of the electrons produced by the $\beta$-decay of free neutrons and supplied by the ionization of the $r$-process elements (the first and second terms), which we ignored in Equation (\ref{tau}). $x$ and $A$ are the degree of ionization and the mean mass number of the $r$-process elements, respectively. Following \citet{Matsumoto2018}, we set $A=80$ and $x=1$ here.

\section{Gravitational lens models}

\begin{figure*}
\centering
\includegraphics[width=0.9\linewidth]{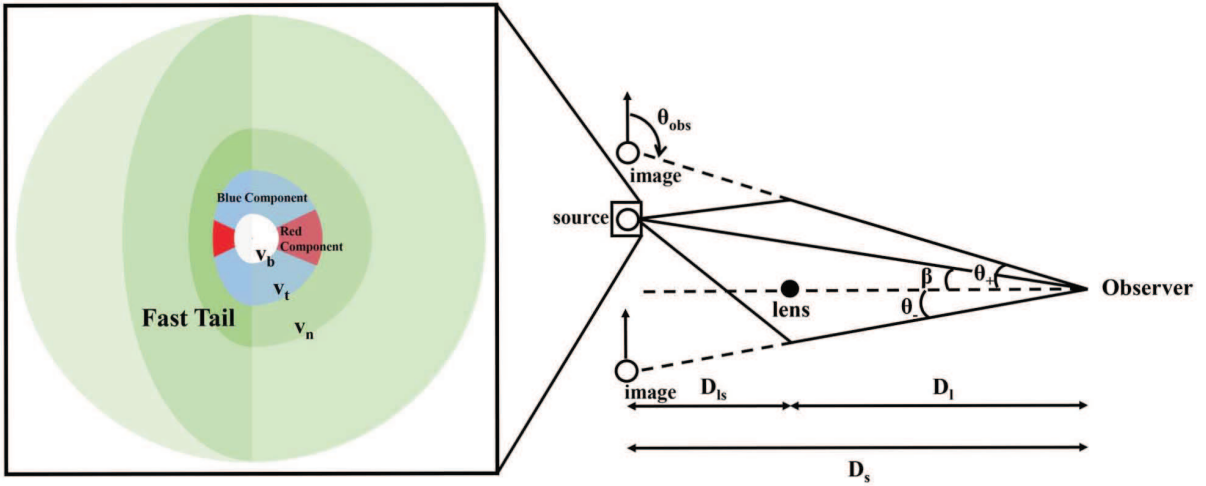}
\caption{Schematic diagrams of kilonovae and the geometry of the PM and the SIS models.}
\label{cartoon}
\end{figure*}

We mainly refer to three mainstream lens models here, including the PM model for treating an object as a point-mass lens, the SIS model for characterizing a galaxy with a particular mass distribution, and the CR model for a PM lens disrupted by a galaxy with an external shear.

\subsection{PM model}

A massive BH with $M_{l}<10^7~M_{\odot}$ can be described as a PM model \citep[][]{Schneider1992}. The Einstein angle that corresponds to the point mass lens $M_{l}$ is defined as
\begin{equation}
\theta_{E}=\sqrt{\frac{4 G M_{l}}{c^{2}} \frac{D_{l s}}{D_{l} D_{s}}}
\end{equation}
where $D_{s}$, $D_{l}$ and $D_{ls}$ are the angular diameter distances to the source at redshift $z_{s}$, to the lens at redshift $z_{l}$, and between the source and the lens, respectively.

The dimensionless lens equation is
\begin{equation}
y=x-\frac{1}{x},
\end{equation}
where $y=\beta/\theta_{\rm E}$, $\beta$ denotes the angular positions of the source, and $x_{\pm}=\theta_{\pm}/\theta_{\rm E}$, where ``+'' and ``$-$'' denote the parity of the image and $\theta_{\pm}$ indicates the angular positions of the image as shown in Figure 1. The corresponding two solutions of the lens equation are
\begin{equation}
x_{\pm}=\frac{1}{2}\left(y \pm \sqrt{y^{2}+4}\right).
\end{equation}

The positive parity image is always the leading image, so the time delay between images $x_{-}$ and $x_{+}$ is
\begin{equation}
\Delta t=\frac{4 G M_{l}}{c^{3}}\left(1+z_{l}\right)\left[\frac{y \sqrt{y^{2}+4}}{2}+\ln \left(\frac{\sqrt{y^{2}+4}+y}{\sqrt{y^{2}+4}-y}\right)\right].
\end{equation}

Their magnifications are
\begin{equation}
\mu_{\pm}=\pm \frac{1}{4}\left[\frac{y}{\sqrt{y^{2}+4}}+\frac{\sqrt{y^{2}+4}}{y} \pm 2\right],
\end{equation}
respectively.

\subsection{SIS model}

The SIS model describes galaxies as gravitational lenses \citep{Schneider1992}. Its Einstein angle can be written as
\begin{equation}
\theta_{E}=\sqrt{\frac{4 G M\left(\theta_{E}\right)}{c^{2}} \frac{D_{l s}}{D_{l} D_{s}}}=4 \pi \frac{\sigma_{v}^{2}}{c^{2}} \frac{D_{l s}}{D_{s}},
\end{equation}
where $M(\theta_{E})$ and $\sigma_{v}$ are the mass within the Einstein radius and the velocity dispersion of the lens, respectively. The corresponding lens equation is
\begin{equation}
y=x-\frac{x}{|x|}.
\end{equation}
If $y<1$, there are two solutions, i.e., $x_{\pm}=y\pm1$. However, if $y>1$, the lens equation only has one solution: $x=y+1$. We only focus on the case of two images to highlight the role of the gravitational lensing effect. The time delay between them is
\begin{equation}
\Delta t=\frac{32 \pi^{2}}{c}\left(\frac{\sigma_{v}}{c}\right)^{4} \frac{D_{l} D_{l s}}{D_{s}}\left(1+z_{l}\right) y.
\end{equation}

The magnifications of the images are
\begin{equation}
\mu_{\pm}=\left|1 \pm \frac{1}{y}\right|.
\end{equation}

\subsection{CR model}

For a point-mass + external shear lens model, i.e., a CR model, an external component of shear from a galaxy is added to the PM lens \citep[e.g.,][]{Chang1979,Chang1984,Schneider1992,Chen2021}. Figure 3 in \citet{Gao2022} depicts the critical curves and transition curves in the plane of the observer at external shear strength $\gamma=0.1$. The observer can perceive four images inside the critical curves and two images beyond this range. Additionally, the later arriving image is brighter inside the range of the transition curves; otherwise, the earlier arriving image is brighter. The dimensionless lens equations are
\begin{equation}
\begin{array}{l}
y_{1}=(1+\gamma) x_{1}-\frac{x_{1}}{x_{1}^{2}+x_{2}^{2}}, \\
y_{2}=(1-\gamma) x_{2}-\frac{x_{2}}{x_{1}^{2}+x_{2}^{2}},
\end{array}
\end{equation}
where we set $\gamma$ equal to $0.1$. The lens equations can have up to four solutions, but they are not easy to derive. Here, we show only special solutions along the axis (i.e., $y_{1}=0$ or $y_{2}=0$).

For $y_{1}=0$, we obtain
\begin{equation}
\begin{array}{l}
x_{1}=0, \\
x_{2}=\frac{y_{2} \pm \sqrt{y_{2}^{2}+4(1-\gamma)}}{2(1-\gamma)};
\end{array}
\end{equation}
and
\begin{equation}\label{y1=0}
\begin{aligned}
x_{1} &=\pm \sqrt{\frac{1}{1+\gamma}-\frac{y_{2}^{2}}{4 \gamma^{2}}}, \\
x_{2} &=-\frac{y_{2}}{2 \gamma}.
\end{aligned}
\end{equation}

For $y_2=0$, we obtain
\begin{equation}
\begin{array}{l}
x_{1}=\frac{y_{1} \pm \sqrt{y_{1}^{2}+4(1+\gamma)}}{2(1+\gamma)}, \\
x_{2}=0;
\end{array}
\end{equation}
and
\begin{equation}\label{y2=0}
\begin{aligned}
x_{1} &=\frac{y_{1}}{2 \gamma}, \\
x_{2} &=\pm \sqrt{\frac{1}{1-\gamma}-\frac{y_{1}^{2}}{4 \gamma^{2}}} .
\end{aligned}
\end{equation}
Note that Equations (\ref{y1=0}) and (\ref{y2=0}) appear only when $\left|y_{2}\right| \leqslant 2 \gamma / \sqrt{1+\gamma}$ and $\left|y_{1}\right| \leqslant 2 \gamma/\sqrt{1-\gamma}$, respectively.
The magnification of each image is
\begin{equation}
\mu=\left(1-\gamma^{2}-\frac{1}{|x|^{4}}-2 \gamma \frac{x_{1}^{2}-x_{2}^{2}}{|x|^{4}}\right)^{-1}.
\end{equation}
The time delay between the two images is
\begin{equation}
\Delta t=t\left(\overrightarrow{x_{s}}\right)-t\left(\overrightarrow{x_{f}}\right),
\end{equation}
where $\overrightarrow{x_{f}}$ and $\overrightarrow{x_{s}}$ are the positions of the earlier and later images, respectively,
and
\begin{equation}
t(\vec{x})=\frac{4 G M_{l}}{c^{3}}\left(1+z_{l}\right)\left[\frac{1}{2}(\vec{x}-\vec{y})^{2}-\ln |\vec{x}|+\frac{\gamma}{2}\left(\vec{x}^{2}-\vec{y}^{2}\right)\right].
\end{equation}

\section{Results}

For various lens models, the associated lensed light curves and the degrees of lensed polarization are derived. Because the lensed images are not resolvable, the kilonova signal we detect is a superposition of signals from different images. The initial time of the first image is used as the starting time. Our observation angles relative to the lensed images are $\theta_{\rm obs}=\frac{\pi}{2}\pm\theta_{\pm}$, respectively, approximately $\theta_{\rm obs}\simeq\frac{\pi}{2}$, because $\theta_{\pm}$ are tiny angles. Figure \ref{cartoon} shows the kilonova component distribution and the geometry of the PM and SIS models. The redshift of the kilonova is kept at $z_{s}=0.01$ in our calculations. The lens redshift can be set to $z_{l}=z_{s}\left(2 z_{s}+3\right) / 3\left(z_{s}+2\right)$ if it is difficult to directly measure the redshift of the lens and to resolve the lens images. This lens redshift selection minimizes the integral of $[M_{l}(z_{l})/M_{l}(z_{\rm true})-1]^{2}$ over $z_{\rm true} = 0-z_{s}$, where $z_{\rm true}$ is the true lens redshift and $z_{\rm true} = 0$ results in the largest possible uncertainty in the lens mass \citep{Chen2021}.

\subsection{Lensed kilonovae light curves}
\begin{figure}
\centering
\includegraphics[width=0.9\linewidth]{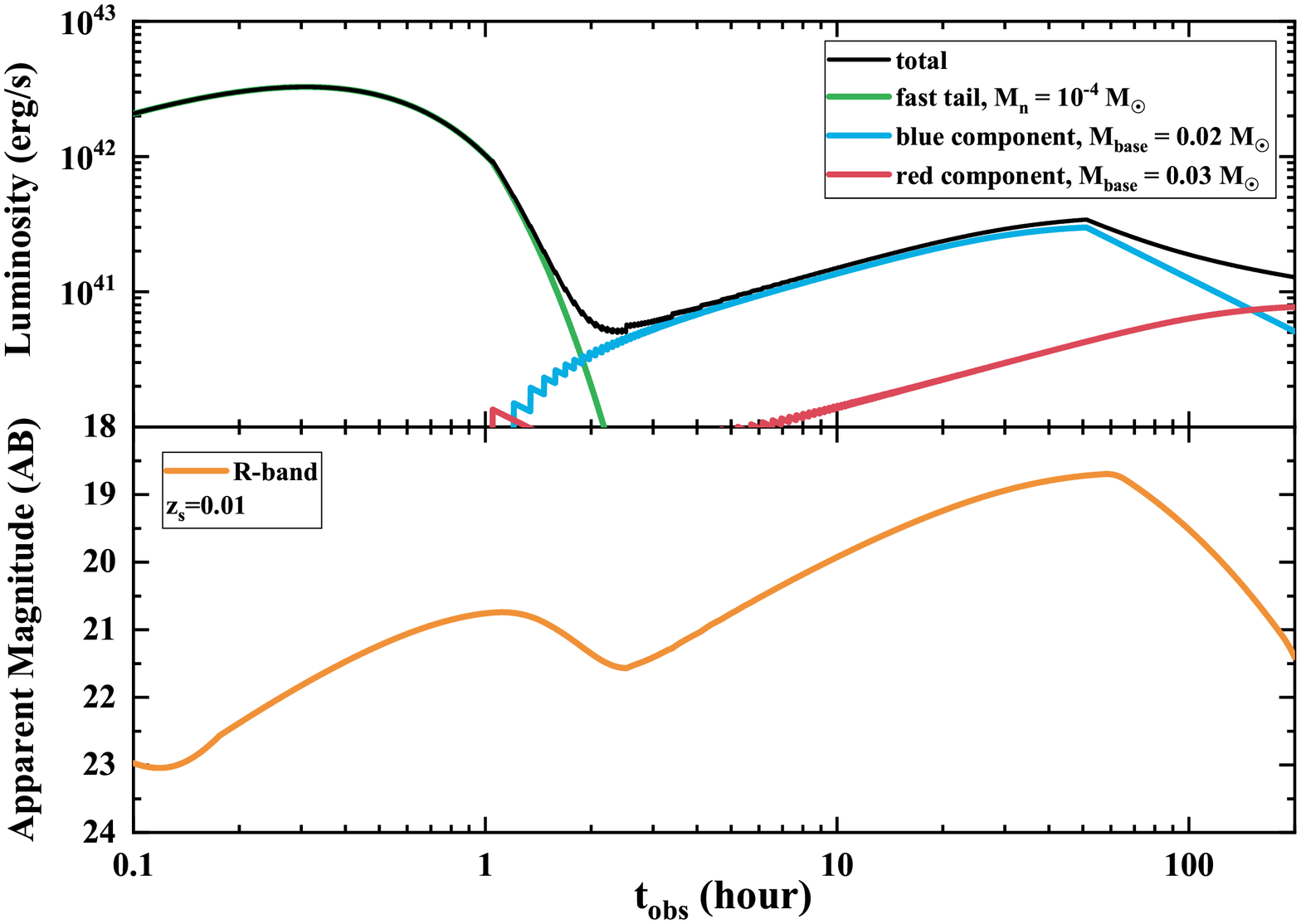}
\caption{Bolometric luminosity (top) and R-band (bottom) light curves of a typical kilonova. A fast tail is assumed to exist outside the red and blue components for kilonovae. Free neutrons of the fast tail undergo $\beta$-decay, producing a precursor at $t_{\rm obs}\sim 1~\rm hr$. The ejecta masses corresponding to the blue and red components, as well as the fast tail, are $0.02$, $0.03$, and $10^{-4}~M_{\odot}$, respectively.}
\label{Light curve}
\end{figure}

The light curves of the bolometric luminosity and apparent magnitude in the R-band without the gravitational lensing effect are shown in Figure \ref{Light curve}. The black total light curve shows that the farthest free neutron heating generates a precursor at $\sim1~\rm hr$, and the peak time of ejecta heating is $\sim60~\rm hrs$. As shown by the orange line, the peak apparent magnitude in the R-band of the ejecta heating is $\sim18.6$.

\begin{figure}
\centering
\includegraphics[width=0.95\linewidth]{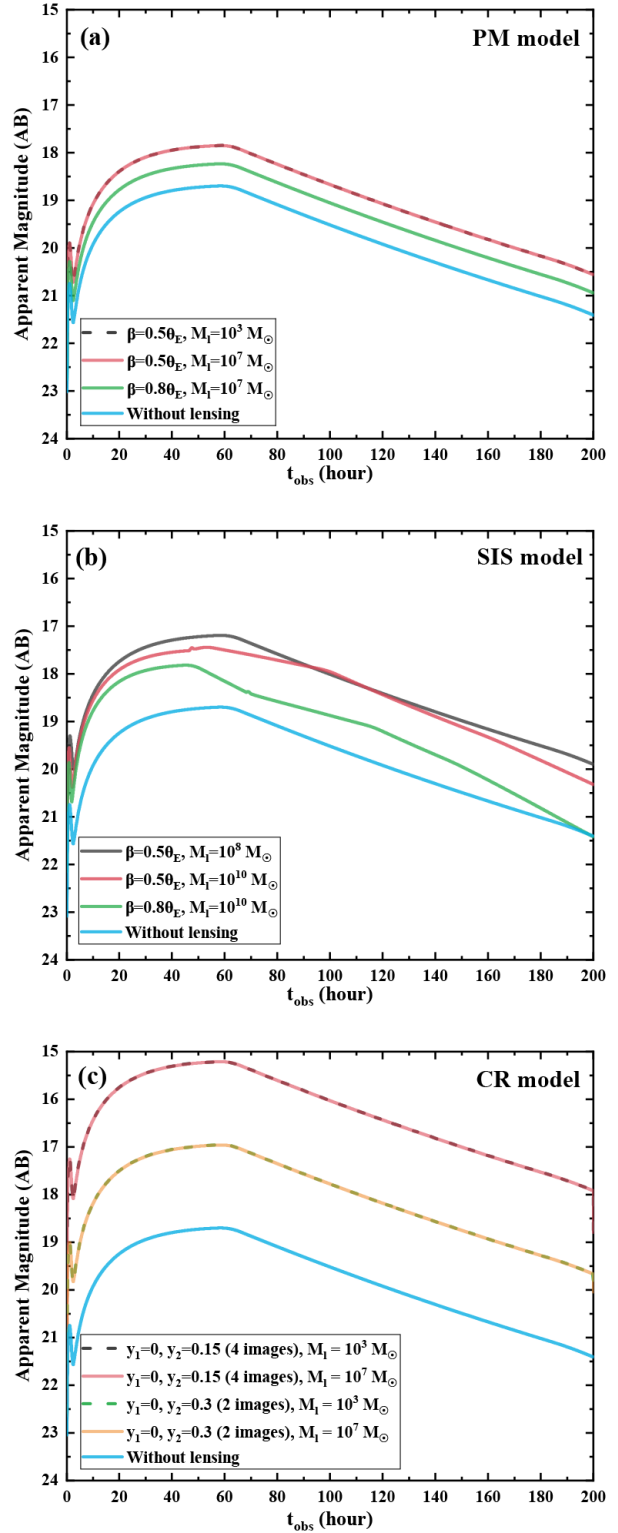}
\caption{Lensed kilonovae light curves in the R-band for the (a) PM, (b) SIS, and (c) CR models with different lens parameters. The solid blue line is the light curve without the gravitational lensing effect.}
\label{Lensed light curves}
\end{figure}

In Figure \ref{Lensed light curves}, to highlight the bump due to the effect of gravitational lensing, we use linear coordinates to prevent confusion with the intrinsic bump of the kilonova in light curves, and the solid blue line indicates the light curve without the gravitational lensing effect. The number of images, the magnification, and the time delay between images all affect the lensed kilonova light curves.

For the PM model, three different cases of lens parameters are configured as follows: ($\beta=0.5~\theta_{E}, M_{l}=10^{3}~M_{\odot}$, black dashed line), ($\beta=0.5~\theta_{E}, M_{l}=10^{7}~M_{\odot}$, red solid line), and ($\beta=0.8~\theta_{E}, M_{l}=10^{7}~M_{\odot}$, green solid line), from which the time delay of two images is on the order of $\sim 10^{2}~\rm s$ or less and the magnification of the second image is smaller than that of the first image. Because the time delay between the two images is significantly less than the ejecta heating timescale, it is impossible to distinguish the flux contribution of each image. Figure \ref{Lensed light curves}(a) depicts a lensed kilonova that was created by superimposing two images with successive arrivals. The magnification of each image is mainly determined by $\beta$, i.e., the source angular position relative to the lens. For the cases of $\beta=0.5~\theta_{\rm E}$ and $\beta=0.8~\theta_{\rm E}$, the peak apparent magnitudes at $\sim60~\rm hrs$ are $\sim17.8$ and $\sim18.2$, and the increase factors are $\sim1.04$ and $\sim1.02$ compared to the case without the gravitational lensing effect (solid blue line), respectively. For the cases of ($\beta=0.5~\theta_{E}, M_{l}=10^{3}~M_{\odot}$) and ($\beta=0.5~\theta_{E}, M_{l}=10^{7}~M_{\odot}$), the light curves are exactly the same due to the same $\beta$ value and the short time delay.

For the SIS model, the lens parameters are ($\beta=0.5~\theta_{E}, M_{l}=10^{8}~M_{\odot}$, black solid line), ($\beta=0.5~\theta_{E}, M_{l}=10^{10}~M_{\odot}$, red solid line), and ($\beta=0.8~\theta_{E}, M_{l}=10^{10}~M_{\odot}$, green solid line). The time delay would be $\sim 10^{3}-10^{5}~\rm s$. As seen in Figure \ref{Lensed light curves}(b), the ($\beta=0.5~\theta_{E}, M_{l}=10^{8}~M_{\odot}$) case has a peak apparent magnitude $\sim17.2$ at $\sim60~\rm hrs$, and it is difficult to distinguish a lens signal due to the short time delay. The ($\beta=0.5~\theta_{E}$, $M_{l}=10^{10}~M_{\odot}$) case has a peak apparent magnitude $\sim17.4$ at $\sim54~\rm hrs$ and shows a lens signal, which is a plateau period with an apparent magnitude of approximately 17.5 between $\sim30$ and $\sim90$ hrs. The ($\beta=0.8~\theta_{E}$, $M_{l}=10^{10}~M_{\odot}$) example also exhibits a tiny lens signal, i.e., an identifiable bump with an apparent magnitude $\sim19$ at $\sim115~\rm hrs$. The reason for the appearance of the lens signal is because the lens mass is large enough to produce a time delay longer than the ejecta heating timescale.

For the CR model, we assumed that the observer's position in the $y_{1}-y_{2}$ plane is $[0, 0.15]$ and $[0, 0.3]$, inside and beyond the range of the critical curve, respectively, so four or two images are available; these positions are within the transition curve range, so the later arriving lens images are brighter than the earlier arriving images. The time delay is on the order of hundreds of seconds and below, so the different images arrive almost simultaneously. In Figure \ref{Lensed light curves}(c), for four images (black and red lines) and two images (green and orange lines), the peak apparent magnitude at $\sim60~\rm hrs$ can reach $\sim15.2$ and $\sim17$, and the increase factors compared to the absence of a lens effect are $\sim1.22$ and $\sim1.09$, respectively. The luminosity of the kilonova can be greatly magnified, but there is no bump-like lensing signal. The CR model has brighter kilonovae than the PM model due to the presence of extra shear from the galaxy.

Based on the lensed light curve results for the PM, SIS, and CR models, we discover that the time delay induced by the SIS model is the most pronounced and results in the most discernible lens signal at $\sim100~\rm hrs$, and that the CR model amplifies the light curve the most. $\beta$ determines the magnification, and the lens mass mainly determines the time delay. When the lens mass is sufficiently large, the time delay is greater than the ejecta heating peak time, and we may directly witness the relatively strong gravitationally lensed kilonova signal.

\subsection{Lensed kilonova polarizations}

\begin{figure}
\centering
\includegraphics[width=0.95\linewidth]{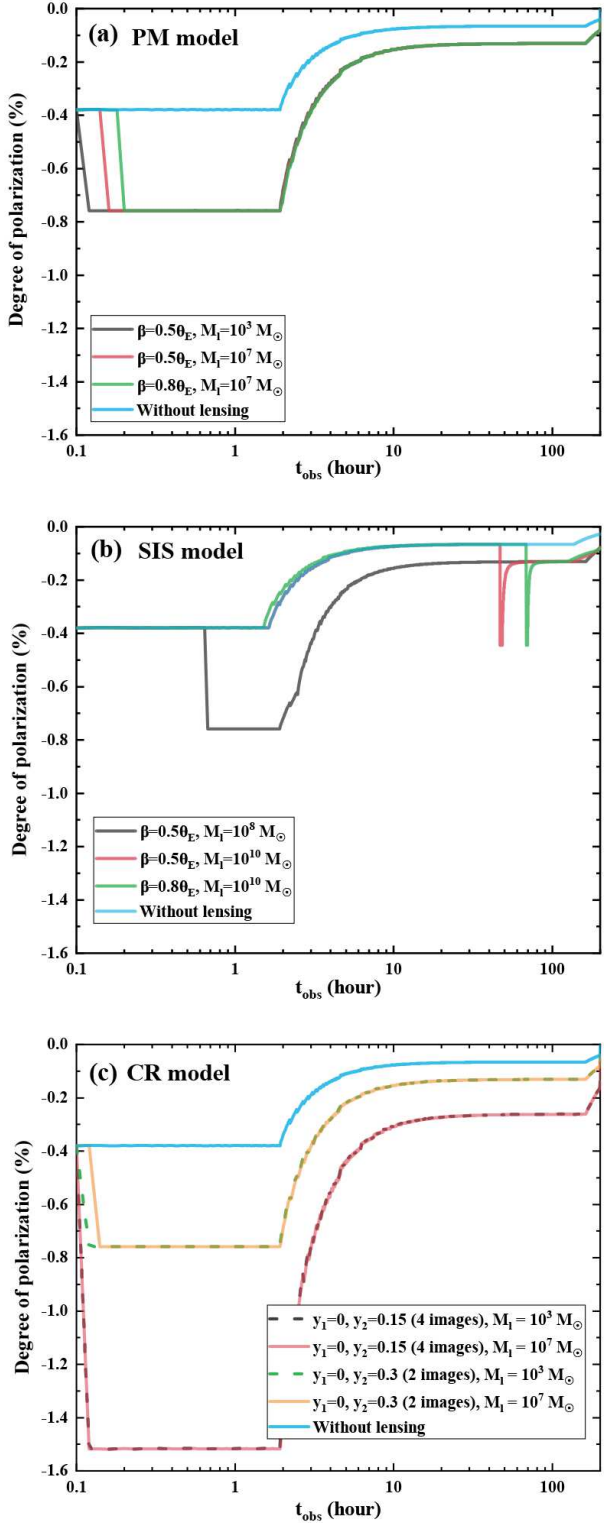}
\caption{Time evolution of the degree of polarization of a lensed kilonova for the (a) PM, (b) SIS, and (c) CR models.}
\label{Lensed polarization}
\end{figure}

The polarization is calculated as a function of electron density by Equation (\ref{polarization}), where polarization in the polar region is negative, and polarization in the equatorial region is positive. We underscore here that the polarization at any given time is the sum of the contributions of all components.

As shown by the blue solid lines in Figure \ref{Lensed polarization}, if there is no gravitational lensing effect, the initial degree of polarization is almost constant with a magnitude of $\sim-0.4\%$ during the first $\sim2~\rm hrs$ and then decreases smoothly. The temporal representation of the polarization can be understood as follows. The photosphere is initially in the layer of free neutrons, $\tau_{\rm es} \simeq 1$, and the total polarization is $\sim-0.4\%$. The timescale of the initial plateau can be used to effectively estimate the total mass of neutrons \citep[e.g.,][]{Matsumoto2018,Li2019}. When the photosphere recedes into the $r$-process element-rich ejecta, the main source of opacity is not electron scattering but bound-bound transition, and the recombination of electrons to the $r$-process elements causes $\tau_{\rm es}$ to drop rapidly, resulting in a rapid decrease in polarization degree. For lensed kilonovae polarization, only the number of images and time delay can affect the degrees.

For the case of the PM model in Figure \ref{Lensed polarization}(a), the polarization with different lens parameters varies mainly in the first 0.2 hrs because the second image is delayed by approximately 100 seconds or less from the first image. When the second image arrives, there is an abrupt rise in the polarization degree. Within the first 2 hrs, the maximum superimposed polarization is approximately $-0.75\%$. Therefore, we can judge from the first 0.2 hrs of polarization whether the kilonova is lensed by the PM lens.

For the SIS model in Figure \ref{Lensed polarization}(b), if $M_{l}=10^{10}~M_{\odot}$, a spike-like bump lasting up to 2 hrs appears at $\sim47~\rm hrs$ and $\sim69~\rm hrs$ with the same polarization degree $-0.45\%$ for $\beta=0.5~\theta_{E}$ and $\beta=0.8~\theta_{E}$, respectively. This is because the time delay between the two images is approximately $10^{3}-10^{5}~\rm s$, and the plateau phase of the second image polarization is overlaid on the falling phase of the first image polarization. For the case of ($\beta=0.5~\theta_{E}, M_{l}=10^{8}~M_{\odot}$, black solid line), the degrees of polarization are $\sim-0.4\%$ between $0.1-0.6~\rm hrs$ and $\sim-0.75\%$ between $0.6-2~\rm hrs$ and then decrease smoothly. If a time delay between two images is greater than 2 hrs, a spike-like bump lasting up to 2 hrs will be produced during the late descending evolutionary stage of polarization, but the maximum polarization will be diminished. If the time delay is less than 2 hrs, a truncated plateau-like elevation will be produced in the first 2 hrs. This can be used to judge whether the kilonova is lensed by the SIS model, and the time delay between the two images can be obtained easily.

For the CR model in Figure \ref{Lensed polarization}(c), the polarization superimposition of four images and two images can reach $\sim-1.52\%$ and $\sim-0.75\%$ within the first 2 hrs, respectively. For the case of four images, the polarization can still be $\sim-0.26\%$ even though the smooth descending stage reaches $100~\rm hrs$.

By comparing the polarization under the three lens models, we find that the time delay and the number of images are the main influencing factors that cause the polarization to vary. The polarization evolution under the PM model mainly differs in the first $0.2~\rm hrs$. For the SIS model, a time delay greater than 2 hrs will result in a truncated bump reaching $\sim-0.75\%$ in the first 2 hrs or a spiky bump lasting up to 2 hrs in the decline phase after 2 hrs. For the CR model, the superposition of the four images makes the total polarization increase greatly. A time delay of more than 2 hrs indicates a reduced maximum polarization.

According to the polarization results of lensed kilonova, the peculiar time variability is the only chance to recognize it. For instance, the PM model shows an abrupt change in polarization at $\sim0.2$ hrs after the merger, but obtaining polarimetry at 0.2 hrs after the merger will be extremely challenging as it requires the kilonova to be unambiguously identified right in the first few minutes after the merger and polarimetric observations to be carried out promptly on a very faint target; and the SIS model shows spike-like bumps lasting up to 2 hrs at $\sim 47$ or $\sim69$ hrs after the merger, the duration of these features are relatively short. So the variation in polarization of lensed kilonovae is relatively difficult to be detected by the current techniques, which typically requires long exposures on faint targets as kilonovae. After accumulating enough samples, it will be possible to distinguish whether there is a lensed kilonova with polarization.

\section{Summary}

Among the GW observation events, only one binary NS merger event has yielded a detected kilonova electromagnetic counterpart thus far. The kilonova ejecta is assumed to consist of $0.03~M_{\odot}$ red and $0.02~M_{\odot}$ blue components, which are mainly present in the equatorial and polar regions, respectively, with a $10^{-4}~M_{\odot}$ fast tail. The heating of free neutrons in the outermost layer of fast tail and $r$-process elements causes the light curve of the kilonova to appear as a precursor at $\sim 1~\rm hr$ and as a peak at $\sim60~\rm hrs$, respectively. In addition, free electrons (products of the $\beta$-decay of neutrons) present in the matter could produce nonzero polarization of $\sim-0.4\%$ within the first 2 hrs. When the free neutron layer subsequently becomes transparent, the degree of polarization diminishes. Of course, the kilonova may be lensed as it propagates to the observer, as in the case of GRBs being gravitationally lensed into multiple images. The PM, SIS, and CR models are three alternative lens models that are considered in this paper while discussing the light curves and polarizations of gravitationally lensed kilonovae.

The lens mass mainly determines the time delay, while $\beta$ and the number of images govern the magnification. According to the results of the lensed light curves, we find that only the lens with $M_{l}\geq10^{10}~M_{\odot}$ could produce a tiny bump-like lens signal at tens of hours in the light curve because the time delay would be $\sim10^{3}-10^{5}~\rm s$ in the SIS model, while the CR model amplifies the peak apparent magnitude up to $\sim15.2$ due to superimposition of four images. When the lens mass is small, as in the PM and CR models, the time delay is on the order of $100\rm~s$ or less, which is less than the ejecta heating timescale, and no substantial lens signal is identified since all images are in the rising phase.

Our analysis of the polarization under the three lens models reveals that the time delay and the number of images are the primary determinants that affect how the polarization varies. The first $\sim0.2~\rm hrs$ of polarization evolution under the PM model differs significantly. A truncated bump reaching $\sim-0.75\%$ in the first 2 hrs or a sharp spike lasting up to 2 hrs in the decline phase after 2 hrs will occur for the SIS model depending on whether the time delay exceeds 2 hrs. The superposition of the four images significantly increases the overall polarization up to 4 times for the CR model. A time delay greater than 2 hrs indicates that the maximum polarization is diminished. \citet{Matsumoto2018} found that if a free nucleon mass of $10^{-5}-10^{-4}~M_{\odot}$ is located within the outermost layer of blue kilonova ejecta from a binary NS merger, the degree of polarisation can be as high as $1-3\%$ for the first $0.3-1~\rm hr$. In the scenario of a BH-NS merger kilonova, \citet{Li2019} demonstrated that if $10^{-4}~M_{\odot}$ of free neutrons persisted in the fastest layer of the dynamical ejecta or disk winds, the degree of polarization during the first hour could be as high as $3\%$ or $0.6\%$, respectively.

Lensed kilonovae are valuable and anticipated transient sources to reveal the nature of compact object mergers, their ejecta morphology, and strong gravitational lenses. Since kilonovae could be brightened in luminosity and heightened in polarization by gravitational lenses, we expect that they might be discovered by future joint multimessenger detections as the samples increase in number.

\acknowledgments
This work was supported by the National Natural Science Foundation of China under grants 12173031 and 12221003.

\end{document}